\documentclass[5p,times]{elsarticle}

\usepackage[numbers]{natbib}
\usepackage{float}
\usepackage{amsmath,amssymb,amsfonts}

\usepackage{subfig}
\usepackage{graphicx}
\usepackage{url}
\usepackage{xcolor}
\usepackage{stfloats}
\usepackage{tcolorbox}
\usepackage[ruled, boxed,linesnumbered]{algorithm2e}
\usepackage{amsmath}
\usepackage{amsfonts}
\usepackage{multirow}
\usepackage{booktabs}
\usepackage{listings}
\usepackage{pifont}
\usepackage{makecell}
\usepackage{tabularx}
\usepackage{array}
\usepackage{listings}
\usepackage{xcolor}
\usepackage{todonotes}
\usepackage{multicol}
\usepackage{booktabs}
\usepackage{multirow}
\usepackage{tcolorbox}
\usepackage{array}
\usepackage{hyperref}
\usepackage{url}
\usepackage{pifont}
\usepackage{tikz}
\usepackage{threeparttable}
\usepackage{titlesec}
\usepackage{enumitem}
\definecolor{verylightgray}{rgb}{.97,.97,.97}

\lstdefinelanguage{Solidity}{
	keywords=[1]{anonymous, assembly, assert, balance, break, call, callcode, case, catch, class, constant, continue, constructor, contract, debugger, default, delegatecall, delete, do, else, emit, event, experimental, export, external, false, finally, for, function, gas, if, implements, import, in, indexed, instanceof, interface, internal, is, length, library, log0, log1, log2, log3, log4, memory, modifier, new, payable, pragma, private, protected, public, pure, push, require, return, returns, revert, selfdestruct, send, solidity, storage, struct, suicide, super, switch, then, this, throw, transfer, true, try, typeof, using, value, view, while, with, addmod, ecrecover, keccak256, mulmod, ripemd160, sha256, sha3}, 
	keywordstyle=[1]\color{blue}\bfseries,
	keywords=[2]{address, bool, byte, bytes, bytes1, bytes2, bytes3, bytes4, bytes5, bytes6, bytes7, bytes8, bytes9, bytes10, bytes11, bytes12, bytes13, bytes14, bytes15, bytes16, bytes17, bytes18, bytes19, bytes20, bytes21, bytes22, bytes23, bytes24, bytes25, bytes26, bytes27, bytes28, bytes29, bytes30, bytes31, bytes32, enum, int, int8, int16, int24, int32, int40, int48, int56, int64, int72, int80, int88, int96, int104, int112, int120, int128, int136, int144, int152, int160, int168, int176, int184, int192, int200, int208, int216, int224, int232, int240, int248, int256, mapping, string, uint, uint8, uint16, uint24, uint32, uint40, uint48, uint56, uint64, uint72, uint80, uint88, uint96, uint104, uint112, uint120, uint128, uint136, uint144, uint152, uint160, uint168, uint176, uint184, uint192, uint200, uint208, uint216, uint224, uint232, uint240, uint248, uint256, var, void, ether, finney, szabo, wei, days, hours, minutes, seconds, weeks, years},	
	keywordstyle=[2]\color{teal}\bfseries,
	keywords=[3]{block, blockhash, coinbase, difficulty, gaslimit, number, timestamp, msg, data, gas, sender, sig, value, now, tx, gasprice, origin},	
	keywordstyle=[3]\color{violet}\bfseries,
	identifierstyle=\color{black},
	sensitive=false,
	comment=[l]{//},
	morecomment=[s]{/*}{*/},
	commentstyle=\color{gray}\ttfamily,
	stringstyle=\color{red}\ttfamily,
	morestring=[b]',
	morestring=[b]"
}

\newtcolorbox{titleEnv}{
colframe=black!80,
colback=gray!10,
fonttitle=\bfseries,
coltitle=black,
left=3pt,
right=3pt,
top=3pt,
bottom=3pt,
boxrule=0.4mm,
arc=3mm
}

\newcommand{\ie}{\textit{i.e.,} }
\newcommand{\eg}{\textit{e.g.,} }
\newcommand{\zyy}{\textcolor{black}}
\newcommand{\zoe}{\color{black}}
\begin{document}

\begin{sloppypar}
\begin{frontmatter}

\title{BinCoFer: Three-Stage Purification for Effective C/C++ Binary Third-Party Library Detection}


\author[HUST]{Yayi Zou}
\ead{yayizou@hust.edu.cn}
	
\author[HUST]{Yixiang Zhang}
\ead{zyx72038@hust.edu.cn}
	
\author[HUST]{Guanghao Zhao}
\ead{petermouse@hust.edu.cn}
	
\author[NTU]{Yueming Wu}
\ead{wuyueming21@gmail.com}

\author[HUST]{Shuhao Shen}
\ead{shuhaoshen@hust.edu.cn}

\author[HUST]{Cai Fu \corref{mycorrespondingauthor}}
\cortext[mycorrespondingauthor]{Corresponding author}
\ead{Fucai@hust.edu.cn}

\address[HUST]{School of Cyber Science and Engineering, Huazhong University of Science and Technology, Wuhan, 430074, China}	
\address[NTU]{School of Computer Science and Engineering, Nanyang Technological University, Singapore}

\begin{abstract}
Third-party libraries (TPL) are becoming increasingly popular to achieve efficient and concise software development. 
However, unregulated use of TPL will introduce legal and security issues in software development.
Consequently, some studies have attempted to detect the reuse of TPLs in target programs by constructing a feature repository. 
Most of the works require access to the source code of TPLs,
while the others suffer from redundancy in the repository, low detection efficiency, and difficulties in detecting partially referenced third-party libraries. 

Therefore, we introduce BinCoFer, a tool designed for detecting TPLs reused in binary programs. We leverage the work of binary code similarity detection(BCSD) to extract binary-format TPL features, making it suitable for scenarios where the source code of TPLs is inaccessible. BinCoFer employs a novel three-stage purification strategy to mitigate feature repository redundancy by highlighting core functions and extracting function-level features, making it applicable to scenarios of partial reuse of TPLs.
We have observed that directly using similarity threshold to determine the reuse between two binary functions is inaccurate, a problem that previous work has not addressed. Thus we design a method that uses weight to aggregate the similarity between functions in the target binary and core functions to ultimately judge the reuse situation with high frequency.
To examine the ability of \emph{BinCoFer}, we compiled a dataset on ArchLinux and conduct comparative experiments on it with other four most related works  (\ie \emph{ModX}, \emph{B2SFinder}, \emph{LibAM} and \emph{BinaryAI}).
Through the experimental results, we find that \emph{BinCoFer} outperforms them by over \zyy{20.0\%} in precision and \zyy{7.0\%} in F1. 
\zyy{As the data volume increases, we observe the precision of BinCoFer tends to be stable and high.}
Moreover, \emph{BinCoFer} greatly accelerates TPL detection efficiency which reduces the time cost of \emph{ModX} by up to 99.7\%.
\end{abstract}


\end{frontmatter}

\section{Introduction} 

With the development of open-source software, modern software is becoming increasingly complex. To improve development efficiency, developers and enterprises often prioritize installing third-party open-source components as dependencies. A significant portion of commercial software is based on open-source code, which is further developed and provided to users in a closed-source form. Verocode's research report \cite{Veracode} indicates that over 96\% of industrial organizations use open-source components in their software application code libraries, and nearly 70\% of applications contain open-source security vulnerabilities. The unregulated usage of third-party libraries (TPLs) can introduce legal issues (\eg license violations \cite{OSSPolice,mancoridis1999bunch,yuan2019B2SFinder,zhan2020automated}) and security problems in software development. For example, accidentally reusing a vulnerable TPL or a TPL containing malware can lead to privacy violations \cite{moonsamy2014android,short2014android,meng2016price,tang2019demystifying,tang2021systematical}, over-privileges, and other security issues.
Although scenarios require the procurement of external software rather than self-development are increasing, providers may not always be able to supply the corresponding Software Bill of Materials (SBOM) files. Moreover, when security issues are exposed in external libraries, the relevant manufacturers may not be able to issue timely alerts or perform necessary repairs.
Monitoring the usage of third-party components in software can better assist in regulating these issues, leading to the development of software composition analysis (SCA) or TPL detection methods.

The general approach to TPL detection involves constructing a TPL feature repository and then performing feature matching on the target software binaries.
We have identified the following key challenges that need to be addressed in the work of software component analysis:

\emph{\textbf{P1: Feature Repository Redundancy and Excess}}.
Functions in one TPL may be copied from another, leading to redundant features being recorded during the construction of the TPL repository, potentially causing false positives in SCA. Additionally, to enhance matching accuracy, some works have extracted a considerable number of features \cite{yang2022modx, li2023libam}, which utilize module/area-based matching. This approach can result in unnecessary space overhead and reduced detection efficiency.

\emph{\textbf{P2: Partially Imported Libraries}}.
When third-party libraries are used, not all functions of the library are referenced. In many cases, especially when libraries are used informally, only key functions are copied. Therefore, there are scenarios where only a part of a third-party library is referenced. If SCA methods rely on the overall features of TPLs, partial reuse may be ignored, reducing recall.

When detecting TPLs in binary form, additional challenges arise:

\emph{\textbf{P3: No Source Code}}.
The versions of TPLs reused in applications may be outdated or restricted by copyright, making it impossible to obtain the source code of the corresponding versions. Consequently, less information can be collected.

\emph{\textbf{P4: Compilation Impact}}.
Binary programs are typically processed with the strip command to remove symbol tables and some debug information, reducing the program size for distribution. Additionally, varying compiler architecture optimization options can result in different binary content from the same source code. This increases the matching difficulty due to potential inconsistencies between the compilation options of the TPL and the binary in the binary-to-binary TPL detection scenario.
Moreover, observations of the BCSD model reveal that the similarity of binary functions obtained from different compilation options of the same source code may be low, while the similarity of different functions may be high(as shown in Figure \ref{fig:funetuned_jtrans}).
Therefore, using a similarity threshold to determine the similarity of two binary code fragments is unreasonable.

Some C/C++ TPL detection works at the binary level have considered these issues, but more research is needed. Classic works \cite{OSSPolice,yuan2019B2SFinder,TPLite,jiang2024binaryai} have utilized features from third-party library source code, which is not feasible when the source code is unavailable. Other works \cite{yang2022modx,li2023libam,tang2022libdb} have designed new matching methods for scenarios involving partial use of third-party libraries, but they have not addressed the redundancy in the feature library, and their matching processes remain resource-intensive.

Recent research on function-level binary code similarity detection (BCSD) aims to alleviate P4 by focusing on robustness across compilers, optimization options, and architectures \cite{massarelli2019safe,xue2018accurate,ding2019asm2vec,liu2018alphadiff,zuo2018neural,redmond2018cross}. Among these, \emph{jTrans} \cite{wang2022jtrans} conducts experiments on a dataset containing 26 million functions with five optimization options, demonstrating that the non-robustness caused by different compilation settings is alleviated. Therefore, it is feasible to consider using function-level features for binary-level TPL detection.

We have summarized three requirements for our method to address the above problems:

\begin{itemize}
    \item \textbf{R1:} Collect the binary form of TPLs to construct the TPL repository, alleviating P3, and select an appropriate purification method to address P1.
    \item \textbf{R2:} Select appropriate granularity and matching methods to detect partially referenced TPLs without using graph matching methods, improving efficiency while ensuring accuracy and recall, thereby addressing P1 and P2.
    \item \textbf{R3:} Address the issue of not directly using a threshold to judge the similarity between two functions in P4, which previous works have not solved.
\end{itemize}

We propose \emph{BinCoFer}, a binary-to-binary C/C++ TPL detection tool to address the aforementioned challenges. For \textbf{R1}, we collect the dynamic library form of TPLs to build the TPL repository. 
\zyy{For \textbf{R2}, we employ our novel three-stage purification process, which, based on the complexity of TPL functions, their reuse within the feature repository, and the principles of library function compilation, to filter out internal functions, simple functions, and reduces the impact of common functions, emphasizing the importance of core functions to eliminate redundancy in the TPL repository.}
To alleviate overhead, we do not extract features from the graph. To detect scenarios where some TPLs are reused, we perform similarity comparisons at the function granularity. To meet \textbf{R3}, instead of setting a threshold to directly determine function reuse, we assign weights \zyy{in the purification process} to highlight core functions and then aggregate the similarity scores of the core functions with the overall similarity score of the target binary to reach a final conclusion.

In summary, our contributions are as follows:
\begin{itemize}

    \item \emph{\textbf{Novel Solution:}} We have designed a solution for scenarios where source code is not available, capable of detecting the partial reuse of TPLs, with a novel three-stage purification technique to highlight the core functions in the C/C++ binary-level TPL feature repository. This alleviates redundancy in the repository and accelerates the detection process. Moreover, the solution addresses the issue of not directly using a threshold to judge the similarity between two functions. We have published both our code and dataset on the same GitHub repository.
    
    \item \emph{\textbf{New Dataset:}} We manually compiled a new ground truth dataset on ArchLinux and published it\footnote{https://github.com/whoami648/BinCoFer}, which includes \zyy{150} binaries and 102 TPLs.

    \item \emph{\textbf{Comprehensive Experiments:}} We conducted experiments on our validation dataset and compared them with previous related works (\ie \emph{ModX} \cite{yang2022modx}, \zyy{\emph{LibAM} \cite{li2023libam}}, \emph{B2SFinder} \cite{yuan2019B2SFinder}, and \emph{BinaryAI} \cite{BinaryAIonline}). Experimental results show that \emph{BinCoFer} outperforms these tools, achieving \zyy{89.3\%} precision and \zyy{64.9\%} recall in detecting 102 TPLs in \zyy{40} manually built binaries. \emph{BinCoFer} beats other TPL detection tools by at least \zyy{20.0\%} in precision and \zyy{7.0\% }in F1 score. Additionally, it reduces the TPL detection time cost of \emph{ModX} by up to 99.7\%.
    
\end{itemize}

\begin{figure*}
    \centering
    \includegraphics[width = 0.8\textwidth]{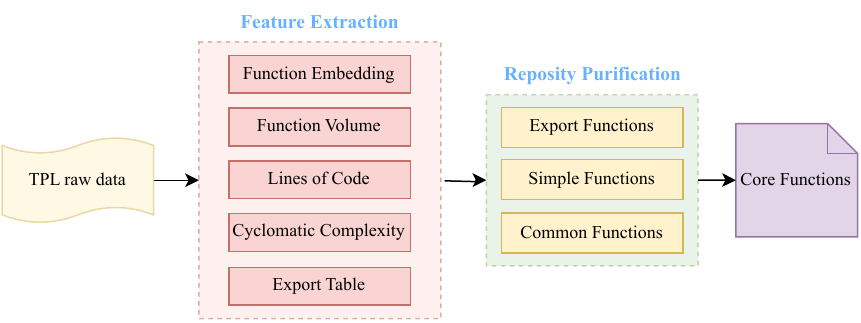}
    \caption{TPL Feature Repository Generation Overview}
    \label{fig:TPL Feature Repository Generation}
\end{figure*}

\textbf{Paper Outline.} Section 2 presents the background, including related work and definitions used in this paper. Section 3 describes the system architecture of \emph{BinCoFer}. Section 4 reports the experimental results. Section 5 discusses the limitations and future work. Section 6 concludes the paper.

\section{Background}

\subsection{Related Work}

In this section, we discuss related works on TPL detection and binary code similarity detection.

\subsubsection{Third Party Library Detection}

Most existing works focus on Java libraries on the Android platform. \emph{LibD} \cite{moonsamy2014androidLibD}, \emph{LibExtractor} \cite{zhang2020empiricalLibExtractor}, and \emph{LibRadar} \cite{ma2016libradar} construct TPL feature repositories using clustering methods based on package structure, homogeneity graphs, and class dependency relations, respectively. These methods can detect TPLs at the library level without prior knowledge. Other studies that do not use clustering methods, such as \emph{LibID} \cite{zhang2019libid}, \emph{LibScout} \cite{demetriou2016freeLibScout}, and \emph{OSSPolice} \cite{OSSPolice}, can accurately detect TPL versions. On the Android platform, according to \cite{zhan2021tplresearch}, more than 70\% of the research targets ad libraries, as these can easily collect user privacy information, leading to privacy leaks. \emph{PEDAL} \cite{liu2015efficientPEDAL} and \emph{AdDetect} \cite{narayanan2014addetect} extract features related to ad libraries to construct feature vectors and use classifiers to distinguish between ad libraries and non-ad libraries.

In practice, there are few studies on TPL detection for native libraries in C/C++. A survey \cite{zhan2021tplresearch} selected 74 relevant works on TPL detection on the Android platform, among which only \emph{OSSPolice} and \emph{NativeGuard} \cite{sun2014nativeguard} support the detection of native libraries. Research on TPL detection in C/C++ source code includes works such as \emph{SourcererCC} \cite{sajnani2016sourcerercc}, which reduces the number of required comparisons for clone detection using token ordering filtering heuristics, proposing an approximate duplicate code detection technology suitable for large code libraries. \emph{CENTRIS} \cite{woo2021centris} proposes a technology that uses redundancy elimination and code segmentation to assist in identifying modified OSS reuse accurately and scalably.

In binary-level TPL detection for C/C++, \emph{BAT} \cite{hemel2011findingBAT} collects string features from the source code of TPLs and assigns lower weights to frequently occurring strings to build a feature library. \emph{B2SFinder} \cite{yuan2019B2SFinder} extracts string, integer, and control-flow features from each TPL, introducing three types of reuse relationships to achieve better performance. However, none of these works propose methods to alleviate the redundancy of the TPL feature database. \emph{OSSPolice} \cite{OSSPolice} utilizes the source file directory structure information and proposes a hierarchical indexing scheme to reduce internal redundancy in the OSS repository, enhancing matching efficiency. \emph{TPLite} \cite{TPLite} proposes a new approach to constructing and verifying TPL dependencies, mitigating the redundancy in the TPL feature library.
However, the aforementioned representative schemes all require the construction of a feature library using the source code of TPLs. When conducting TPL detection, features are extracted at the granularity of the entire TPL, making it difficult to handle complex situations where the TPLs are partially reused.

In the binary-to-binary TPL detection scenario, \emph{OSLDetector} \cite{zhang2020osldetector} directly extracts features from the binary, generating weights and building an internal clone forest to filter features and alleviate the redundancy of the feature library. However, it does not use semantically relevant features and has a relatively small test dataset. \emph{ModX} \cite{yang2022modx} proposes a modular division approach to address the partial reuse of TPL, but the efficiency of both module division and similarity matching between modules is low, leading to excessive time consumption. It also does not offer a solution to address the redundancy of the feature library. \emph{LibAM} \cite{li2023libam} refines the detection granularity into areas with stronger interpretability than modules for partial reuse of TPL, applying the detection scenario to a compilation environment across optimized option architectures, but the similarity matching between areas is also time-consuming.

\subsubsection{Binary Code Similarity Detection}

Research on binary code similarity detection (BCSD) aims to detect the degree of similarity between two code fragments and has important applications in vulnerability detection, malware detection \cite{bruschi2006detecting,cesare2013control,chandramohan2016bingo}, and software supply chain analysis, such as software component analysis and reverse engineering.

Existing BCSD works extract features at the granularity of basic blocks, functions, and programs, using static and dynamic methods to compare two binary code fragments based on syntax, semantics, and structure. Researchers further study the robustness of BCSD in scenarios such as cross-architecture, cross-compiler, and code obfuscation. Dynamic methods such as \emph{BLEX} \cite{egele2014blanketBLEX} and \emph{IMF} \cite{wang2017memoryIMF} execute program fragments in a controlled simulation environment and judge the similarity of functions based on observed runtime behavior features. However, they heavily depend on the architecture and have low practicality.

Initial research in static methods used manually extracted features, such as operator or instruction sequence features in \emph{BinClone} \cite{farhadi2014binclone}, \emph{ILine} \cite{jang2013towardsILine}, \emph{MutantX-S} \cite{hu2013mutantx}, \emph{BinSign} \cite{nouh2017binsign}, and \emph{Kam1n0} \cite{ding2016kam1n0}, and graph features in \emph{TEDEM} \cite{pewny2014leveragingTEDEM} and \emph{XMATCH} \cite{feng2017extractingXMATCH}. However, these methods tend to overlook the semantic information of instruction or basic block sequences.

Later, machine learning was combined with feature extraction to automatically map internal features of functions to low-dimensional vectors for comparison. Some works such as \emph{Gemini} \cite{xu2017neuralGemini}, \emph{VulSeeker} \cite{gao2018vulseeker}, and \emph{GraphEmb} \cite{massarelli2019investigatingGraphEmb} use graph neural networks (GNN) to generate graph embeddings of attributed control-flow graphs (ACFG) of binary functions, performing similarity comparison at the graph level. Other works such as \emph{SAFE} \zyy{\cite{Massarelli2018SAFESF}} and \emph{ASM2VEC} \cite{ding2019asm2vec} are based on natural language processing (NLP), and some works such as \emph{BAR} \zyy{\cite{Massarelli2019InvestigatingGE}} and \emph{OrderMatters} \cite{yu2020Ordermatters} combine GNN and NLP. The recent work \emph{jTrans} \cite{wang2022jtrans} uses transformer-based \cite{vaswani2017attention} language models to learn representations of binary code and achieves state-of-the-art results on large-scale datasets for cross-compilation tasks. We fine-tune \emph{jTrans}'s model to generate function-granularity feature embeddings to construct the TPL feature repository.

\subsection{Definition}

\subsubsection{Third Party Library}

To simplify the development process and improve the scalability and reusability of programs, many reusable code modules in programs on systems such as Windows and Linux are written by other individuals or organizations. These reusable code modules are called third-party libraries (TPLs). These libraries provide common functionalities that programmers can use to quickly develop software without having to write all the code from scratch. However, due to the lack of systematic supervision, inappropriate use of TPLs may lead to a series of legal disputes and security issues, making the task of detecting TPL reuse essential.

\subsubsection{Core Function\zyy{s}}\label{sec:Core Function}

Core functions refer to a collection of functions that represent the core algorithmic logic of a code library. We believe that some of the core functions must be reused when a TPL is reused. The characteristics of these functions are unique to each library. Ideally, we assume that there is no overlap between the core functions of different TPLs. However, the implementation is limited by existing binary code similarity technology, making it impossible to directly filter and retain the core functions. Therefore, when detecting TPL, we assign higher weights to more core functions.
\subsubsection{Simple Function\zyy{s}}

\begin{figure}
\centering

\subfloat[Function$_1$:free\_AttributeType]{
\begin{minipage}[t]{0.2\textwidth}
\centering

\includegraphics[width=\textwidth]{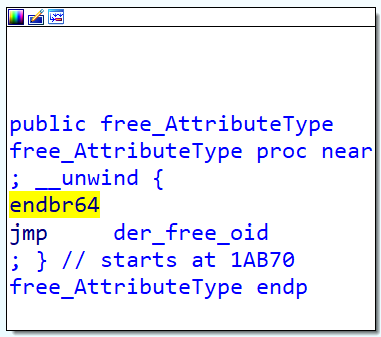}
\end{minipage}
}
\hspace{0.03\textwidth} 
\subfloat[Function$_2$:length\_AttributeValue]{
\begin{minipage}[t]{0.22\textwidth}
\centering
\includegraphics[width=\textwidth]{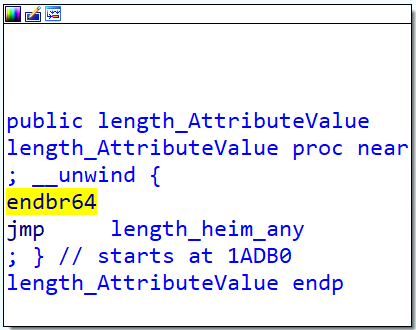}
\end{minipage}
}
\centering
\caption{Control Flow Graph (CFG) of two functions in libasn1.so.8.0.0}
\label{fig:functions}
\end{figure}

Simple functions refer to functions that do not contain any core logical algorithms. Retaining these functions in the feature repository would occupy storage resources. Such functions generally have simple code logic, small size, and are not complex. Figure \ref{fig:functions} shows an extreme case where the control flow graph (CFG) of the function has only one basic block, containing no logical assembly code other than a jump statement.

\subsubsection{Common Functions}\label{Common Function}

Common functions refer to functions with a high frequency of occurrence and those pointed to by frequently occurring features in the feature repository. Such functions are not unique enough to represent the core functionality of a specific library, and their weights need to be reduced.

Since the binary code may change if the program source is recompiled with different compilation processes, even when the source code does not change, the \zyy{`}frequency of function occurrence' in our paper does not refer to the \zyy{`}number of function embeddings that are exactly the same as the feature embedding of this function' in the repository, but rather to the \zyy{`}number of embeddings that are similar to it'. The actual meaning of \zyy{`}common functions' in this scenario can have several interpretations.

First, there is a code clone phenomenon among TPLs, where some commonly cloned functions have a high frequency of occurrence in the feature repository. These cloned functions are classified as \zyy{`}common functions', resulting in the embedding of a function appearing in different TPLs.

Second, functions implementing common functionalities that are not unique to a library are also classified as \zyy{`}common functions'. None of the libraries in the TPL repository are the original ones implementing these common functions. The algorithmic functionalities implemented in common functions can be easily reproduced by different TPLs, but the feature embeddings generated by different implementations of the same algorithm at the binary level may vary significantly, making it difficult to directly determine the similarity of function embeddings. Therefore, \zyy{`}reproduced algorithms' in our work only refers to functions implemented similarly at the source code level. Such function feature embeddings can appear in different TPLs.

Third, some structurally simple functions can be similar to other functions within a TPL or among TPLs after normalization or other preprocessing operations. This phenomenon often occurs when the disassembled function body contains only a single \zyy{`}jmp' statement, referencing an external function or other functions within the TPL's \zyy{`}.text' section. For example, Figure \ref{fig:functions} shows the control flow graph (CFG) of two functions, \emph{free\_AttributeType} and \emph{length\_AttributeValue}, from the \zyy{`}.text' section of the \zyy{`}libasn1.so.8.0.0' library. Each function contains only one statement that jumps to another function in the \zyy{`}.text' section, resulting in identical function embeddings after normalization.

\section{Methodology}

\begin{figure*}
    \centering
    \includegraphics[width = 0.95\textwidth]{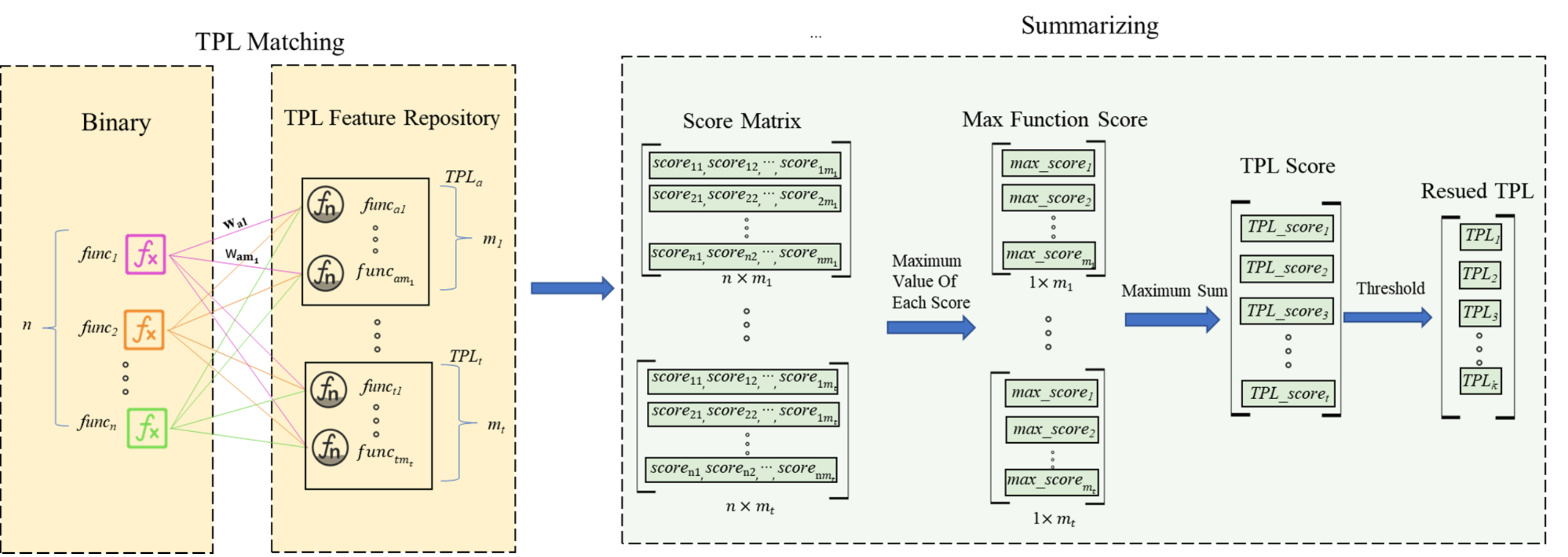}
    \caption{TPL Detection Overview}
    \label{fig:TPL Detection}
\end{figure*}

\subsection{Overview}

Figures \ref{fig:TPL Feature Repository Generation} and \ref{fig:TPL Detection} illustrate the overall architecture of \emph{BinCoFer}, which consists of two stages: \emph{TPL feature repository generation} and \emph{TPL detection}.

In the first stage, concepts from source code are migrated to binary code to design some features of TPLs, using function-level BCSD technology to construct the TPL feature repository. Critical functions within the TPL feature repository are identified and assigned higher matching weights. In the second stage, \emph{BinCoFer} directly compares the similarity of function embeddings between the target binary and the TPL feature repository, ultimately summarizing the similarity scores to judge the relationship between the binary and each TPL.

\subsection{TPL Feature Repository Generation}

The TPL feature repository generation, as shown in Figure \ref{fig:TPL Feature Repository Generation}, consists of two parts: feature extraction and repository purification. The purpose of feature extraction is to construct a TPL feature repository at the function granularity for detection and to extract features that determine the core functions.
The repository purification stage filters the TPL feature repository based on the extracted function importance features by setting a threshold to remove non-core functions and assign importance weights to core functions in the feature library.

\subsubsection{Feature Extraction}\label{Feature Extraction}

The features of third-party libraries refer to code attribute features and importance features at the function level. All features of all TPL functions are retained during the TPL feature repository extraction stage. The aim is to collect data, extract features of function code attributes, and identify features of function importance to highlight core functions. 
Before commencing feature extraction, we need to construct our dataset. We selected the top 100 most frequently used packages in the ArchLinux system, as well as the TPLs involved in the ground truth dataset we compiled. We filtered out some TPLs that were inaccurately decompiled by IDA, ultimately obtaining a candidate library consisting of 102 TPLs.

\textbf{1) Code Attribute Features Generation.} We adopted the definition of binary code similarity from the survey \cite{haq2021survey}, which states that if two pieces of binary code share syntactic, structural, or semantic similarity, they can be considered similar. We use the features that determine the similarity between two code segments as code attribute features.
It should be noted that we use \zyy{`}similar score' instead of \zyy{`}identical' to determine the reuse of TPL key functions because the compilation environment of dynamic link libraries often differs from the binary, and binaries compiled from the same source code may differ due to different compiler options and architectures.

\begin{figure}[ht]
  \centering 
  \subfloat[same function pairs similarity]{%
    \begin{minipage}[t]{0.4\textwidth}
      \includegraphics[width=\textwidth]{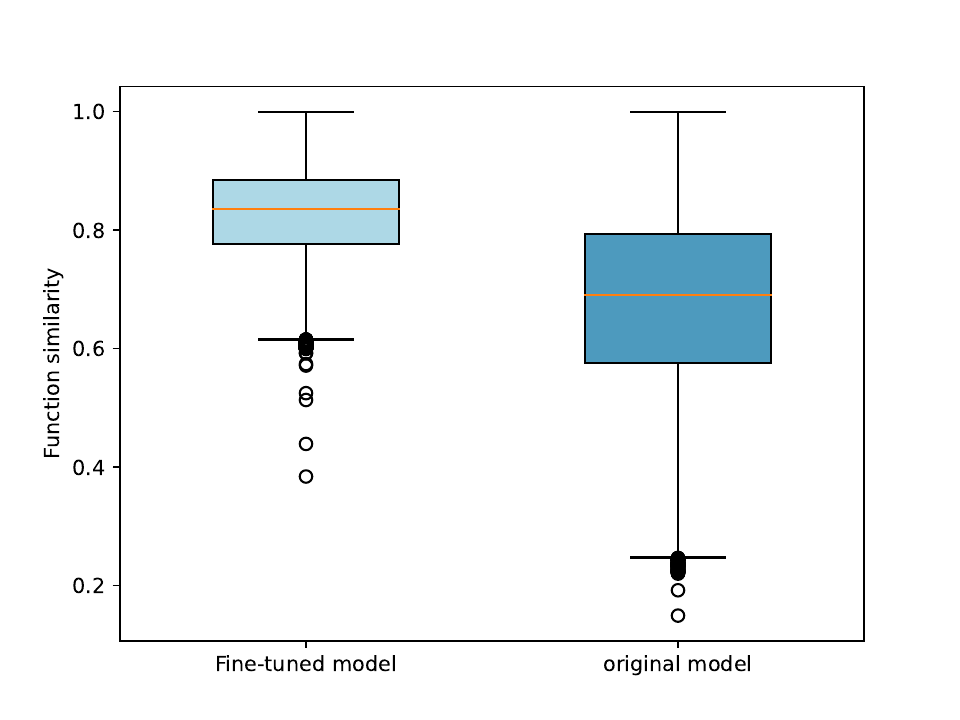}
    \end{minipage}%
    \label{fig:sub1}
  }
  \hfill 
  \subfloat[different function pairs similarity]{%
    \begin{minipage}[t]{0.4\textwidth}
      \includegraphics[width=\textwidth]{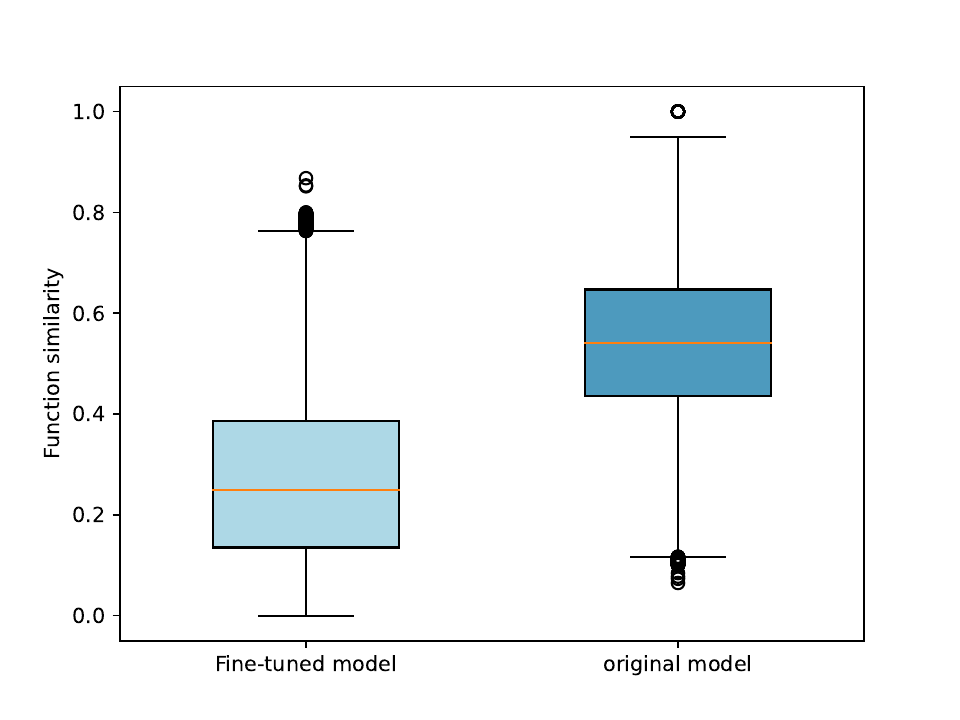}
    \end{minipage}%
    \label{fig:sub2}
  }
  \caption{Similarity Distribution of same and different function pairs detected by the models before and after fine-tuning with jTrans}
  \label{fig:funetuned_jtrans}
\end{figure}

To effectively extract binary function code attribute features for similarity comparison, we chose the \emph{jTrans} \cite{wang2022jtrans} method, which uses a transformer model to extract semantic embeddings of functions. It combines assembly instruction information and jump structure information to achieve SOTA results in function similarity detection on large-scale datasets with different compilation options. We fine-tuned the jTrans model on the dataset from \emph{LibAM} \cite{li2023libam} to distinguish same and different pairs of functions by similarity. Figure \ref{fig:funetuned_jtrans} shows the model's performance before and after fine-tuning.


We use the generated embedding representation as the code attribute feature. The outputs of this step are the function embeddings of all functions.

\textbf{2) Importance Features Generation.} Importance features are a significant basis for determining whether a function is a core function. Programs are composed of functions, which consist of assembly instructions. A function can be regarded as a control flow graph composed of multiple basic blocks, and different functions in a program may be connected through a function call graph. We extract importance features based on the structural information within and outside of these function bodies as well as other external information.
Our work does not aim to match the binary with a specific version of TPL precisely, so we only collect one version of each TPL when building the TPL feature repository. We preprocess the collected data with the decompilation tool \emph{IDA Pro 7.5} \cite{ida}, and then extract the following information from the metadata.

\textbf{Feature 1: Function Volume.} In programs, larger functions often perform core functionality, while smaller functions may be utility functions. Therefore, function size is one of the features used to measure the importance of a function. The Halstead complexity metrics, developed by Maurice Halstead, is a quantitative measure of program complexity at the source code level, determined from the operators and operands in modules. Halstead's volume (HV) describes the size of the algorithm implementation, calculated using Equation (\ref{HV1}).
\begin{equation}
HV = N*log_2(n)\label{HV1}
\end{equation}
\begin{equation}
N = N_{1}+N_{2}\label{HV2}
\end{equation}
\begin{equation}\label{HV3}
n = n_{1}+n_{2}
\end{equation}
The program length (N) is the sum of the total number of operators (N1) and operands (N2) in the program. The vocabulary size (n) is the sum of the number of unique operators (n1) and operands (n2). We designed to migrate HV to binary, mapping the operators and operands in the source code to the operation codes and operands in binary assembly instructions, and directly applying the HV Equation \ref{HV1} to calculate the function volume. The larger the HV, the more complex the algorithm and the more likely it is to include core functions.

\textbf{Feature 2: Lines Of Code.} In binary programs, the number of lines of function code is defined as the number of function assembly instructions excluding blank lines. This is a high-level supplementary measure of function volume and one of the features for evaluating the importance of functions.

\textbf{Feature 3: Cyclomatic Complexity.} Cyclomatic complexity \cite{mccabe1976complexitycc} is a measure of the complexity of a module's decision-making structure. Mathematically, the cyclomatic complexity of a structured program is defined with reference to the program's control flow graph and is calculated using Equation \ref{CC1}.
\begin{equation}
CC = E-N+2P\label{CC1}
\end{equation}
E is the number of edges in the control flow graph. N is the number of nodes in the graph. P is the number of connected components. For a single program (or subroutine or method), P is always equal to 1, simplifying the formula to Equation \ref{CC2}.
\begin{equation}
CC = E-N+2\label{CC2}
\end{equation}
Since control flow graphs also exist in binary functions, we apply the calculation method of cyclomatic complexity to measure the complexity of binary functions. E represents the number of edges in the function's control flow graph, and N represents the number of nodes (basic blocks) in the control flow graph. The higher the cyclomatic complexity, the greater the program's complexity, often implying more logical branches and decision points. We use this feature to measure the importance of functions.

\textbf{Feature 4: Export Table.} According to the authoritative guide to IDA Pro, the export window shows the entry point of program execution (specified in the program's file header and named \zyy{`}start' in IDA Pro) as well as any functions and variables exported from the file for use in other files. We believe that functions in a TPL that provide external functionality are more representative of its key features, often defined in the source code with keywords such as \zyy{`}extern' or \zyy{`}attribute'. Therefore, in our experiments, we used IDA to extract the functions from the export table of the dynamic link library, thus filtering out other irrelevant functions in the TPL feature repository.

\subsubsection{Repository Purification}\label{sec: Repository Purification}

In the second stage of building the TPL feature repository, we purify the repository based on the features generated in the first stage that can be used to evaluate the importance of functions. We retain export functions, remove simple functions, and reduce the importance of common functions. The repository selection includes three purification steps, as shown in Figure \ref{fig:TPL Feature Repository Generation}.

\textbf{1) Keep Export Functions.} 
In the Feature Extraction Stage \ref{Feature Extraction}, we labeled each function in the TPL that comes from the export table as an export function. The TPLs we collected were stored in the form of dynamic link libraries. According to the characteristics of dynamic link libraries, not all functions can be used by other modules; functions are only allowed to be called externally after being exported. Therefore, we believe that only the functions that provide external interfaces for TPL can represent its core functionality. However, not all functions in the export table can necessarily be considered core functions of the TPL. Based on our observations, some functions in the export table, as read by the disassembly tool IDA Pro, do not have any additional code written in them, and their function body only contains a single statement that directly jumps to an external function. For such functions, we filter them out based on their function complexity in the later section. Therefore, in this part, we only use the export table to filter out internal functions, thus retaining more functions that represent the core functionality of the TPL exposed to other files.

\textbf{2) Simple Functions Filtering.} 
Simple functions generally exhibit simple code logic, small size, and lack of complexity. We calculate the number of assembly code lines at a high level and borrow the concept of cyclomatic complexity \cite{mccabe1976complexitycc} from the source code to measure structural complexity within the function body. Additionally, we borrow the concept of Halstead Volume from the source code to measure the size of a function from the perspective of operator and operand levels. Finally, we combine these features and calculate the overall complexity of the function using the maintainability index (MI), as shown in Equation \ref{MI}:
\begin{equation}\label{MI}
MI = 171-5.2*\ln(HV)-0.23*CC-16.2*\ln(LOC)
\end{equation}
Here, HV represents Halstead Volume, CC represents cyclomatic complexity, and LOC represents the count of lines of code. These are all extracted in the feature extraction stage, and the details for calculating HV and CC are shown in Equations \ref{HV1} and \ref{CC2}. The maintainability index \cite{oman1992definitionMI} is a software metric that measures how maintainable (easy to support and change) the source code is, with higher values indicating better maintainability. We borrowed the complexity calculation formula from the source code and applied it to binary assembly code to measure the complexity of binary functions in this work. We use a percentage threshold to filter out simple functions with small size and little complexity (higher MI).

We set a filter percentage threshold rather than a threshold based on the frequency of occurrence because the setting of the threshold requires empirical statistics on large-scale datasets. Since we only conducted experiments on a relatively small TPL feature repository, we use a filter percentage threshold to make our method more adaptable to changes in the dataset and more scalable. It should be noted that we did not modify the coefficients in the MI calculation formula directly from the source code, because the MI complexity function we ultimately obtained is not used to directly determine whether a function is complex, but to measure the relative complexity between different functions. Therefore, the parameters in Equation \ref{MI} can be roughly set without much impact on the final effect. However, if we want to more accurately measure the maintainability of the source code from the perspective of binary programs, more statistical work is needed.

\textbf{3) Common Functions Marginalization.} 
According to the definition of common functions in section \ref{Common Function}, functions with a high frequency of occurrence in the repository do not typically represent core functionalities in most programs. We aim to reduce the impact of such functions in TPLs that merely reuse these functions. However, since we can only obtain the TPLs in binary form and do not have access to related information such as the creation time of functions in each TPL, it is impossible for us to determine which clone functions originate from. Additionally, we cannot directly use thresholds to decide the similarity between two functions. Therefore, the frequency of occurrence is not sufficiently precise for us to directly use it to filter out common functions. Instead, we use a TF-IDF-based approach as shown in Equation \ref{TFIDF} to assign lower weights to more common functions.

\begin{align}\label{TFIDF}
	\begin{split}
		&TF(f,l) = \frac{n_{f,l}}{|l|}, \\
		&IDF(f)=\log{\frac{|L|}{df_{f}+1}},\\
            &TF-IDF(f,l)={TF(f,l)}\times{IDF(f)}
	\end{split}
\end{align}

For a given function $f$ and a third-party library $l$, $n_{f,l}$ represents the number of occurrences of function $f$ within the third-party library $l$. $|l|$ stands for the total number of functions in a third-party library. $df_{f}$ is the count of how many third-party libraries reuse function $f$ in the repository. $|L|$ signifies the total number of third-party libraries in the repository. The value of $TF-IDF(f,l)$ is the product of the $TF(f,l)$ value and the $IDF(f)$ value.
\subsection{TPL Detection}

After completing the two stages of feature repository generation, we highlight the core functions for each TPL in the repository. As shown in Figure \ref{fig:TPL Detection}, during TPL detection, we extract and embed features for all functions in the target binary and then compare them one by one with the features in the TPL feature repository.
According to the observations in Figure \ref{fig:funetuned_jtrans}, it is not accurate to use the threshold directly to determine that two functions are similar, so we retain the highest similarity score for each core function within one TPL when compared to the functions in the target binary. By applying weights to these scores, we summarize the similarity score between the target binary and each TPL. Finally, a threshold \zyy{$\theta_3$} is established to judge whether the binary reuses a TPL.
\zyy{
\begin{align}\label{overall_score}
	\begin{split}
		&\text{Score}_{ki\_aj} = \text{TF-IDF}(f_{aj}, a) \times \cos(\mathbf{e}(f_{ki}), \mathbf{e}(f_{aj})), \\
		&\text{max\_score}_{ki\_a} = \max_{j}(\text{Score}_{ki\_aj}),\\
            &\text{Score}_{ka} = \sum_{i=1}^{n} \text{max\_score}_{ki\_a}
	\end{split}
\end{align}
We present the specific algorithm in  Equation \ref{overall_score},
The score between the \( i \)-th function in the \( k \)-th binary and the \( j \)-th function in the \( a \)-th TPL is denoted by \( \text{Score}_{ki\_aj} \). The maximum similarity score between the \( i \)-th function in the \( k \)-th binary and any function in the \( a \)-th TPL is represented by \( \text{max\_score}_{ki\_a} \). Finally, we sum up the \( n \) similarity scores between the \( k \)-th binary and the \( a \)-th TPL is denoted by \( \text{Score}_{ka} \).
Here, \( n \) represents the number of functions in the binary.
\text{If } \( \text{Score}_{ki\_aj} \) > $\theta_3$, \text{ then the } k\text{-th binary reuses the } a\text{-th TPL.}
}
\section{Evaluation}
In this section, we construct a dataset on the ArchLinux platform and conduct experiments. We select appropriate threshold combinations in Section \ref{sec:Threshold Selection(RQ1)}. Based on these thresholds, we further evaluate the effectiveness (RQ2) and scalability (RQ4) of \emph{BinCoFer} and test the impact of each filtering stage (RQ3). The research questions are as follows:

\begin{itemize}
    \item \emph{\textbf{RQ1: }What do several thresholds mean in the experiment, and how do we select them?}
    \item \emph{\textbf{RQ2: }What is the precision and recall of \emph{BinCoFer} compared to related works in detecting TPLs?}
    \item \emph{\textbf{RQ3: }What is the impact of each filtering stage on the final TPL detection performance under the selected threshold?}
    \item \emph{\textbf{RQ4: }How is the scalability of \emph{BinCoFer} in terms of \zyy{data volume and its} time efficiency \zyy{ when} compared to related works?}
    
\end{itemize}

\subsection{Evaluation Setup}

\subsubsection{Comparative Systems}

There are few TPL detection works for C/C++. In our experiment, we selected the following four benchmarks for comparison:

\emph{\textbf{ModX}} \cite{yang2022modx} is one of the latest binary-to-binary TPL detection works for C/C++, using a community detection approach to partition the binary program to be detected into modules (a collection of functions that form a complete set of functionality) and comparing them with the modules in the TPL feature repository based on similarity.

\emph{\textbf{LibAM}} \cite{li2023libam} is a novel area matching framework that connects isolated functions into function areas based on FCG and then detects TPLs by comparing the similarity of function areas. It mitigates the impact of different optimization options and architectures and provides interpretable evidence for TPL detection results by identifying exact reuse areas.

\emph{\textbf{B2SFinder}} \cite{yuan2019B2SFinder} is one of the most advanced binary-to-source SCA works, extracting strings, values, and structures for each TPL at file granularity and introducing three new concepts of reuse types. We migrated it to the binary-to-binary scenario for comparison, selecting the four original features suitable for binary as described in Libdb \cite{tang2022libdb}.

\emph{\textbf{BinaryAI}} \cite{BinaryAIonline} is an online software component analysis platform launched by Tencent KeenLab. Their latest paper \cite{jiang2024binaryai} shows that they decompile binary programs with Ghidra to Pcode and train a source-Pcode model for TPL detection.

\subsubsection{Dataset}\label{sec: dataset}

Since some datasets used in previous works are small or not entirely open source, and it is challenging to collect identical data directly from a given list, we decided to compile a new dataset ourselves.
We initially attempted to compile projects on GitHub with distinct \zyy{`}third-party\zyy{'} folders to construct the ground truth. However, we found that there was essentially no reuse of TPLs between these third-party projects, making it unsuitable for addressing the feature repository redundancy scenario discussed in this paper.
The dataset exposed by LibAM \cite{li2023libam} was also compiled from GitHub projects and may have the same properties.
Considering that the BCSD model we use is trained on a dataset from ArchLinux, we chose to compile programs from ArchLinux to allow the BCSD model to detect on a similarly distributed dataset.

We scraped all packages in the ArchLinux system as of August 2023, totaling 11,369 projects. By manually modifying the PKGBUILD documents, we added static linking compilation options and adjusted the compilation order to change dynamic linking to static linking, ultimately compiling \zyy{150} binary programs.
\zyy{In principle, a static library is composed of multiple `.o' files. A binary program compiled through static linking will only include the `.o' file containing the function that it has reused, rather than including all functions from the entire library. Therefore, the compiled dataset mimics the scenario of partial reuse.}

To select a TPL repository closer to real-world scenarios, we added the top 100 popular TPLs used in the ArchLinux system. After filtering out libraries with IDA decompilation errors, we obtained a TPL repository consisting of 102 TPLs.

We divided the dataset using a 7:3 ratio, with \zyy{73\%} (110 binaries) used to select the threshold (threshold selection dataset) and \zyy{27\%} (\zyy{40} binaries) used to validate the effect under the selected threshold (validation dataset).
We did not change the compile optimization options in the PKGBUILD document, so most of the resulting binary compilation options are O2, and a few are O0, included in both the training and testing sets.

\subsubsection{Metrics}

We adopt widely used metrics to measure performance. The descriptions of our used metrics are as follows:

\begin{itemize}
    \item \emph{True Positive} (TP): the number of libraries correctly detected as reused libraries.
    \item \emph{False Positive} (FP): the number of libraries incorrectly detected as reused libraries.
    \item \emph{False Negative} (FN): the number of libraries incorrectly detected as unreused libraries.
    \item \emph{Precision} = TP / (TP + FP). The correct rate of detection.
    \item \emph{Recall} = TP / (TP + FN). The percentage of clone pairs successfully detected.
    \item \emph{F1} = 2$*$Precision$*$Recall / (Precision + Recall). A comprehensive metric of detection.
\end{itemize}

\subsection{Threshold Selection (RQ1)} \label{sec:Threshold Selection(RQ1)}

When filtering out functions that retain core functionality, we set three thresholds. The first is the similarity threshold $\theta_1$, which determines the likelihood that two function embeddings are similar using cosine similarity but not accurate enough to directly determine whether two functions came from the same source code.
The second is the percentage threshold $\theta_2$, which filters out simple functions based on the percentage of retained functions.
The third is the similarity threshold $\theta_3$, which measures the similarity between the binary and TPL.
\zyy{Since the similarity between two functions cannot directly represent the degree of similarity between them when comparing at the function level, in most cases, only a rough distinction can be made based on the overall function similarity ranking. Therefore, we tested different threshold combinations.}

\begin{figure}
    \centering
    \includegraphics[width = 0.4\textwidth]{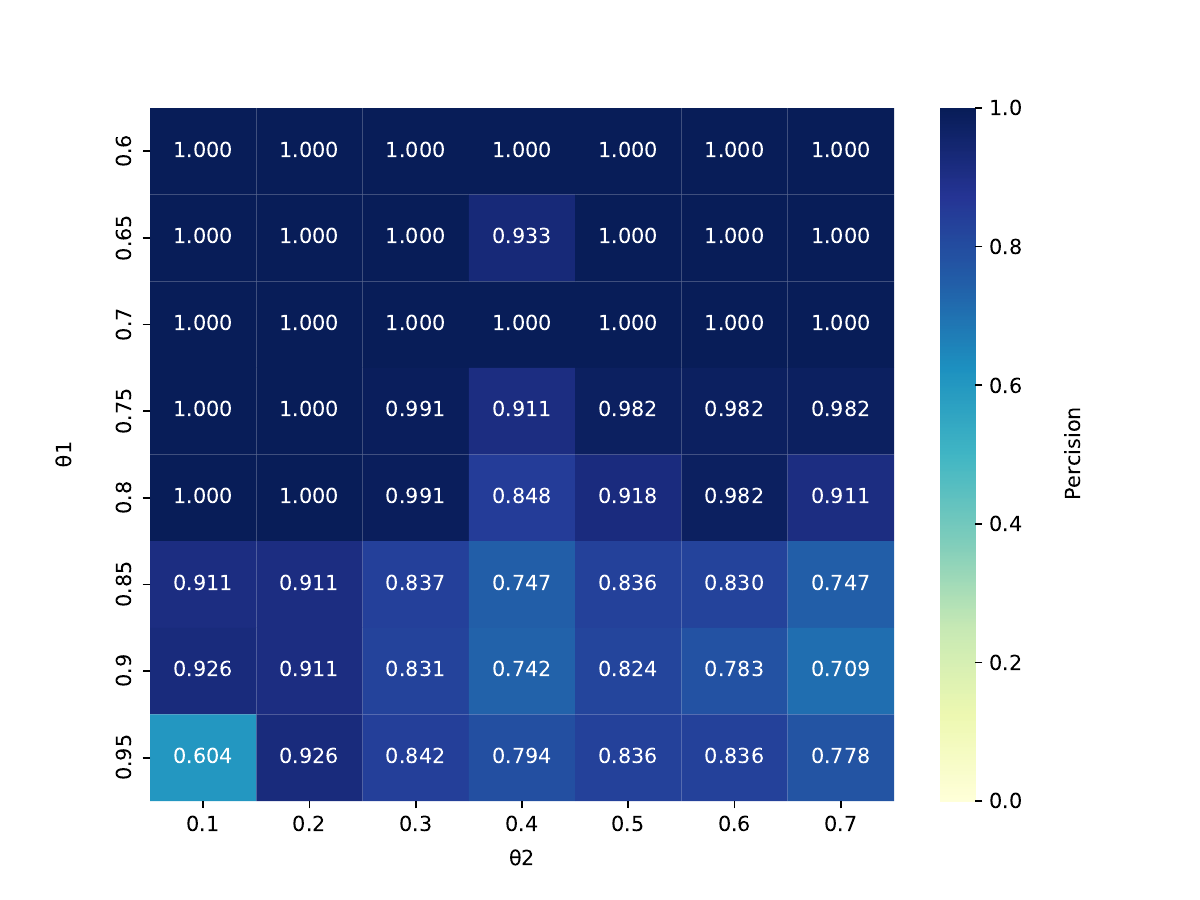}
    \caption{The heatmap of Precision as $\theta_1$ and $\theta_2$ vary with fixed values of $\theta_3$ = 0.89}
    \label{fig:heatmap_Percision}
\end{figure}

\begin{figure}
    \centering
    \includegraphics[width = 0.4\textwidth]{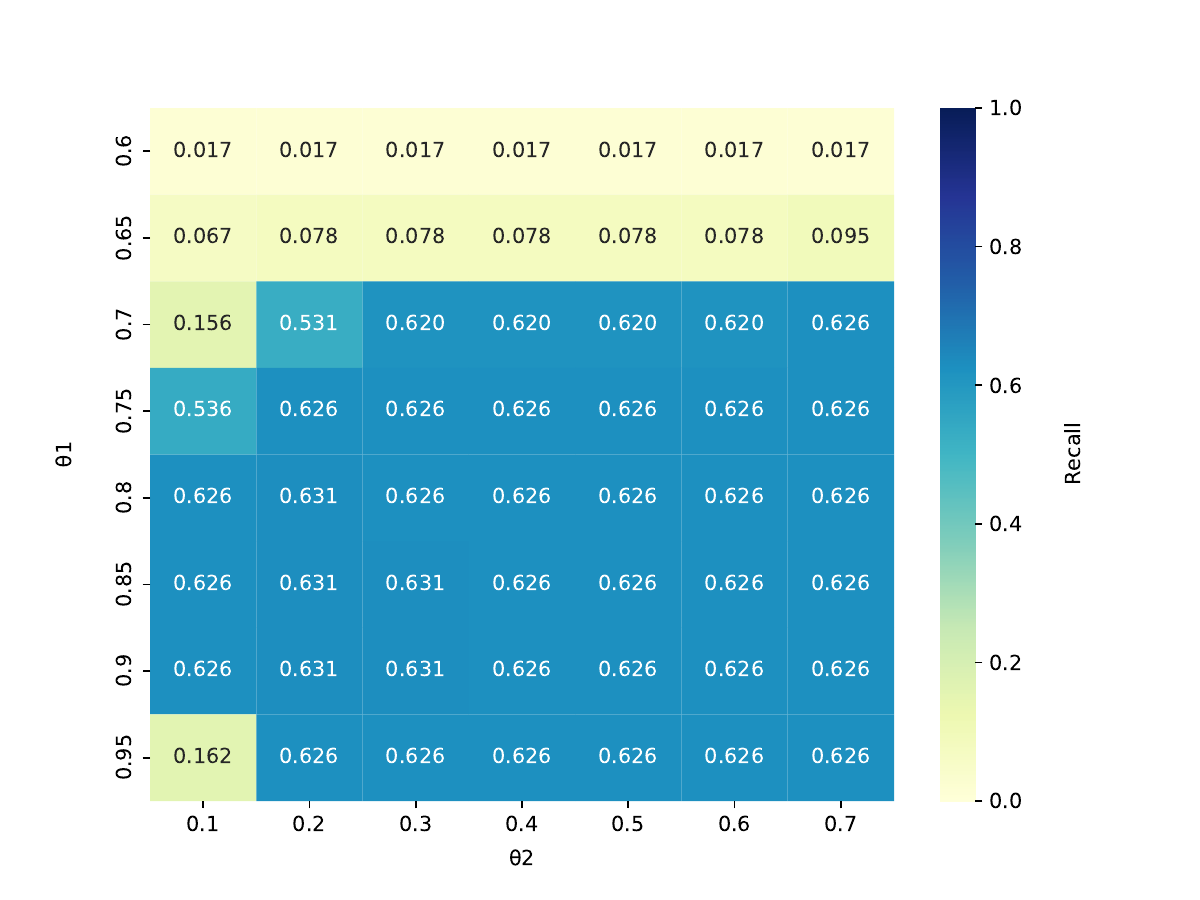}
    \caption{The heatmap of Recall as $\theta_1$ and $\theta_2$ vary with fixed values of $\theta_3$ = 0.89}
    \label{fig:heatmap_Recall}
\end{figure}

\begin{figure}
    \centering
    \includegraphics[width = 0.4\textwidth]{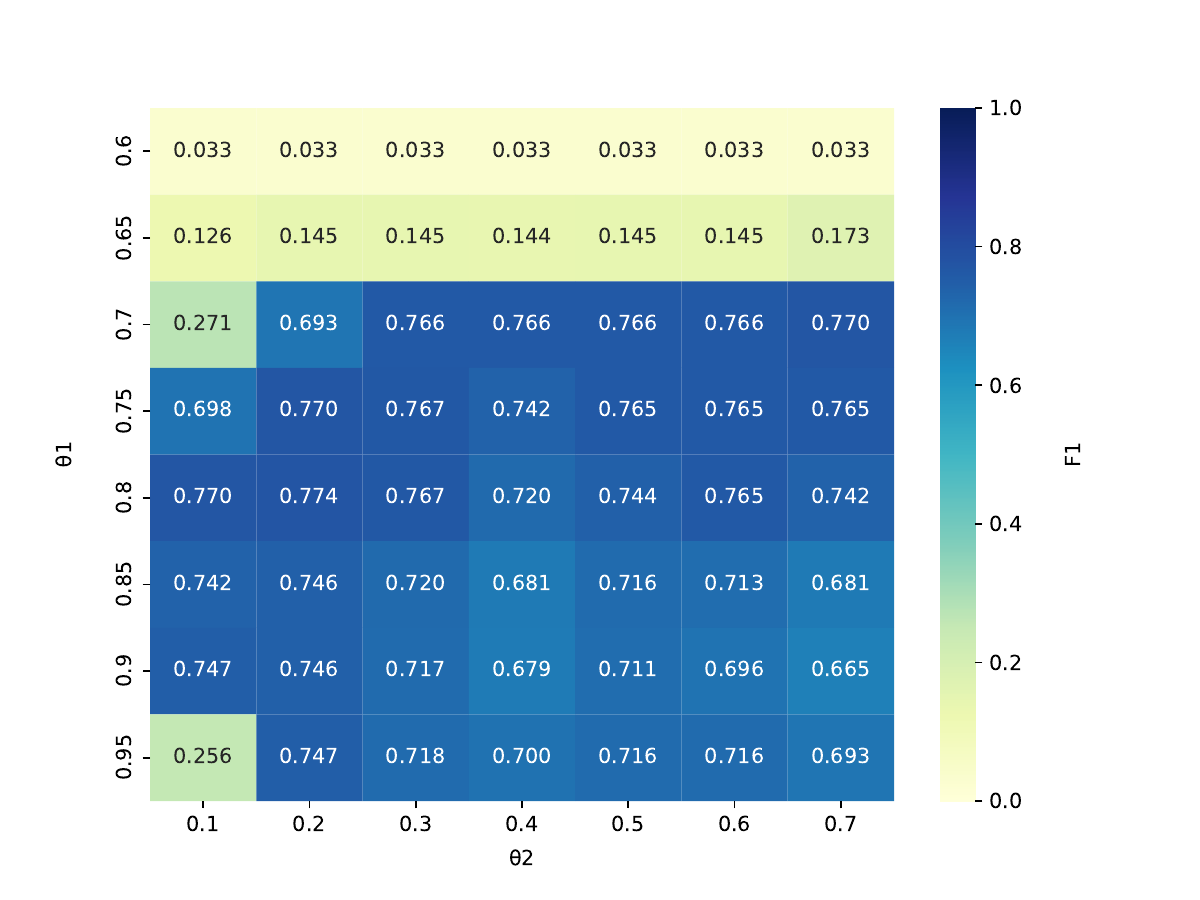}
    \caption{The heatmap of F1 as $\theta_1$ and $\theta_2$ vary with fixed values of $\theta_3$ = 0.89}
    \label{fig:heatmap_F1}
\end{figure}

We began by fine-tuning the original fine-tuned model published by Jtrans \zyy{with its original fine-tuning method} using data from LibAM \cite{li2023libam} as shown in Figure \ref{fig:funetuned_jtrans}. We \zyy{selected} the threshold range of $\theta_1$ using the binary code similarity task on the BinaryCorp small\_test dataset \cite{wang2022jtrans}, where the \zyy{fine-tuned Jtrans} model achieved the highest F1.
\zyy{Thus we set the range of $\theta_1$ in range(0.75, 0.95, 0.05).
Then we set the ranges for $\theta_2$ and $\theta_3$ as $\theta_2$ in range(0.1, 0.7, 0.1) and $\theta_3$ in range(0.7, 0.95, 0.01) based on observations (after setting a broader observation range).}
We performed threshold selection operations on our threshold selection dataset, observing heatmaps of Precision, Recall, and F1 as shown in Figures \ref{fig:heatmap_Percision}, \ref{fig:heatmap_Recall}, and \ref{fig:heatmap_F1} when fixing $\theta_3$ at 0.89.

We primarily chose based on F1, selecting the threshold combination with the highest F1 and more functions preserved in the TPL feature repository: $\theta_1$ = 0.8, $\theta_2$ = 0.2, $\theta_3$ = 0.89.

\begin{titleEnv}
\textit{\textbf{Answering RQ1:}
We set three thresholds: $\theta_1$ (measuring the similarity between functions), $\theta_2$ (the percentage of functions to be preserved based on their MI complexity), and $\theta_3$ (measuring the similarity between binary and TPL). When adjusting the thresholds, we select the combination that maximizes F1 \zyy{of BinCoFer in threshold selecting dataset} while preserving more functions \zyy{in TPL repository}. We select $\theta_1$ = 0.8, $\theta_2$ = 0.2, and $\theta_3$ = 0.89.}
\end{titleEnv}

\subsection{Accuracy Evaluation (RQ2)} \label{sec:Accuracy Evaluation(RQ2)}

Based on the selected thresholds, we compared the precision and recall on our \zyy{hold-out} validation dataset (\zyy{40} binaries) and compared the results of \emph{BinCoFer} with \emph{ModX} \cite{yang2022modx}, \emph{B2SFinder} \cite{yuan2019B2SFinder}, \emph{LibAM} \cite{li2023libam}, and \emph{BinaryAI} \cite{BinaryAIonline}. We reproduced the entire experiment of ModX with the help of its authors. We ran LibAM from the source code available on GitHub. Since B2SFinder is designed for binary-to-source SCA scenarios, we only extracted four features applicable to the binary-to-binary case,
as described in Libdb \cite{tang2022libdb}. We examined \zyy{the same 40} binaries using BinaryAI's latest release (2024.2.26). The final results are shown in Table \ref{tab:Performance Summary}.

\begin{table}[htbp]
\small
\setlength{\tabcolsep}{0.12cm}

\caption{Performance Comparison}
\begin{tabular}{|c|ccccc|}
\hline
 & BinCoFer & B2SFinder & ModX & LibAM & BinaryAI \\
\hline
Precision & \zyy{0.893} & \zyy{0.679} & \zyy{0.693} & \zyy{0.384} & \zyy{0.684} \\
Recall & \zyy{0.649} & \zyy{0.609} & \zyy{0.648} & \zyy{0.936} & \zyy{0.684}\\
F1 & \zyy{0.752} & \zyy{0.642} & \zyy{0.669} & \zyy{0.545} & \zyy{0.684} \\
\hline
\end{tabular}
\label{tab:Performance Summary}
\end{table}

We can see that when performing TPL detection on each binary using \emph{BinCoFer}, the detection results achieved a precision of \zyy{89.3\%} and a recall of \zyy{64.9\%}. We believe this is the result of setting appropriate thresholds, heavily filtering non-core functions in the TPL feature repository.
\zyy{In fact, we did not deliberately choose scenarios with the highest precision, instead, we selected the threshold combination based on the highest F1 score on the sub-datasets, and coincidentally, the precision at this point happened to be relatively higher. This phenomenon may also suggest that our approach achieves the best overall performance in scenarios where precision is given more emphasis.}
Since numerical values of function occurrence frequency and MI complexity cannot strictly determine whether a function is a core function, a few core functions will inevitably be filtered out, resulting in some \zyy{FNs}. However, the results still prove the effectiveness of the proposed scheme against the redundancy of the TPL repository.

\zyy{By comparison, both of the recall rate and precision of the other works are close except for LibAM.
We found that the high FP rate of LibAM leads to its low precision. We use LibAM to observe the reuse of TPL inside the repository and find that some TPLs in the FPs reused the TPLs from the ground truth, confirming the impact of TPL reuse in the repository and the importance of purification.
This also explains the phenomenon of poor performance of ModX.
Comparatively, LibAM is more suitable for scenarios with higher requirements for recall.
ModX uses modular division and is implemented for detecting partially reused third-party libraries and assumes no reuse between TPLs in the repository, leading to more FPs in scenarios involving TPL reuse.
BinaryAI uses its own TPL repository for detection, and the redundancy of the repository may not be removed, resulting in certain FPs. Both ModX and BinaryAI rely on deep learning models, and their models may not have seen data on ArchLinux, leading to inaccuracies.
B2SFinder uses simple code features, does not rely on deep learning models, and does not have a semantic understanding, so it tends to generate a large number of FPs and FNs when detecting more challenging cases.
}

\begin{titleEnv}
\textit{\textbf{Answering RQ2:}
BinCoFer achieves \zyy{89.3\%} accuracy and \zyy{64.9\%} recall on our own compiled ArchLinux validation dataset of \zyy{40} binaries, \zyy{at least 20\%} improvement in precision and \zyy{7\% improvement in F1 over B2SFinder, ModX, LibAM, and BinaryAI, which verifies the importance of the purification step.}
}
\end{titleEnv}

\subsection{Ablation Experiment (RQ3)} \label{sec:Ablation Experiment(RQ3)}

To verify the effectiveness of the filters at each step, we conducted ablation experiments based on the thresholds $\theta_1$ = 0.80, $\theta_2$ = 0.2, and $\theta_3$ = 0.89 selected in RQ1.

We assigned higher weights to core functions in a TPL in a TF-IDF-like manner in the experiment shown in Table \ref{tab:Ablation Experiment with weight}, and gave equal weight to every function in a TPL in the experiment shown in Table \ref{tab:Ablation Experiment without weight}.
The headers in both tables have the same meaning. As shown in Table \ref{tab:Ablation Experiment with weight}, we compared the TPL detection results at different stages of filtering the feature repository\zyy{,} results at different stages of filtering the feature repository. \zyy{`}Origin' refers to the number of functions extracted at the beginning with functions in \zyy{`}.plt', \zyy{`}extern', \zyy{`}.init', and \zyy{`}.fini' segments filtered out during extraction.The functions in these sections are unrelated to the actual implementation of the functions in the library. All subsequent experiments were conducted based on the 123,867 functions in Origin. \zyy{`}Export' refers to retaining export functions. \zyy{`}MI' refers to filtering simple functions based on their MI complexity. \zyy{`}Export+MI' refers to performing the \zyy{`}MI' operation after \zyy{`}Export'.
In the ablation experiments, we combined these filters in different orders, where the order of combination indicates that the subsequent filter is based on the results of the previous one.
For example, in Table \ref{tab:Ablation Experiment with weight}, after the \zyy{`}Export' filter, the percentage of remaining functions in the TPL repository is 0.468. Since $\theta_2$ = 0.2, the percentage of remaining functions after the \zyy{`}Export+MI' filter is 0.092. Although $\theta_2$ = 0.2, the actual percentage of retention is less than 0.2 because the corresponding MI complexity threshold is calculated based on this percentage threshold. Many functions in the dataset have the same MI complexity, and we filtered functions with an MI complexity equal to the threshold, leading to a lower retention percentage. The precision, recall, and F1 of each experiment are shown in Table \ref{tab:Ablation Experiment with weight}.
According to the experimental results in Table \ref{tab:Ablation Experiment with weight}, each individual filter operation improves precision while decreasing recall.
\begin{table}[htbp]
\small
\setlength{\tabcolsep}{0.19cm}
\caption{Ablation Experiment without weight}
\renewcommand{\arraystretch}{1.2} %
\begin{tabular}{|c|cccc|}
\hline
        & Orgin & Export & MI & Export+MI\\ 
\hline
        Func Num (\#) & 123,867 & 57,999 & 23,432 & 11,499 \\ 
        Leave Percent & 1.000  & 0.468  & 0.189  & 0.092 \\ 
\hline
        Percision & \zyy{0.048}  & \zyy{0.218}  & \zyy{0.376}  & \zyy{0.526} \\
        Recall & 1.000  & \zyy{0.706}  & \zyy{0.698} & \zyy{0.652} \\ 
        F1 & \zyy{0.092}  & \zyy{0.333}  & \zyy{0.488}  & \zyy{0.582}  \\ 
\hline
\end{tabular}
\label{tab:Ablation Experiment without weight}
\end{table}

\begin{table}[htbp]
\small
\setlength{\tabcolsep}{0.19cm}
\caption{Ablation Experiment with weight}
\renewcommand{\arraystretch}{1.2} %
\begin{tabular}{|c|cccc|}
\hline
        & Orgin & Export & MI & Export+MI\\
\hline
        Func Num (\#) & 123,867 & 57,999 & 23,432 & 11,499 \\
        Leave Percent & 1.000 & 0.468 & 0.189 & 0.092 \\
\hline
        Percision & \zyy{0.098} & \zyy{0.224} & \zyy{0.469} & \zyy{0.893} \\ 
        Recall & \zyy{0.747} & \zyy{0.656} & \zyy{0.678} & \zyy{0.649} \\
        F1 & \zyy{0.173} & \zyy{0.333} & \zyy{0.554} & \zyy{0.751} \\
\hline
\end{tabular}
\label{tab:Ablation Experiment with weight}
\end{table}

\zyy{As shown in Table \ref{tab:Ablation Experiment without weight}, the recall is 1.0 when experiments are performed on `origin' without weight, shows that the function similarity threshold and binary similarity threshold are reasonable.}
\zyy{Comparing the experimental results from Tables \ref{tab:Ablation Experiment with weight} and \ref{tab:Ablation Experiment without weight}, assigning higher weights to core functions has improved the final overall experimental results, which shows the solution is effective. 
However, the overall recall in Table \ref{tab:Ablation Experiment with weight} is lower than that in Table \ref{tab:Ablation Experiment without weight}.}
The result verifies the second limitation point, because the common functions are not known to belong to, and they are given lower matching weights, some binary checks that reuse these common functions will produce missing alarms.
\zyy{When filtering the feature repository with only the export table and test with weight, the precision rises and the recall rate drops, which indicate that TPL repository originally contained internal functions to interfere with detection.
However, in theory, removing the internal functions of TPL should not affect the detection results. The possible reasons for this could be that setting a function similarity threshold can inevitably lead to FNs and FPs to some extent, or there might be issues with the disassembling tool IDA.}
Using \zyy{`}MI' alone to filter the feature repository \zyy{and test with weight} also significantly improves precision, \zyy{more than 80\% of the functions in the repository are filtered out,} with the recall rate only dropping by \zyy{0.07} while the precision rate increased by \zyy{0.37}. \zyy{Which indicates that there} are simple functions in the dataset, causing significant interference during detection.
Ultimately, using \zyy{`}Export+MI' \zyy{and test with weight}, over 90\% of the functions were filtered out. With the recall rate only decreasing by \zyy{0.098 compared with `origin'}, the precision rate achieved a \zyy{much higher} score of \zyy{0.893}, validating the effectiveness of the approach presented in this paper for scenarios involving redundancy in the feature library.

\begin{titleEnv}
\textit{\textbf{Answering RQ3:}
Based on the results of the ablation experiments, each individual filtering operation improved precision with only a small diminishing effect on recall. 
After filtering the TPL repository with the export table and MI complexity, giving more weights to core functions in a TPL in a TF-IDF-like manner achieved the best experimental result. It is speculated that simple functions in the TPL feature repository have a greater disturbance on TPL detection.
}
\end{titleEnv}

\subsection{Scalability Evaluation (RQ4)} \label{sec:Scalabiliy Evaluation(RQ4)}

Regarding scalability, we first measured the number of features in the feature repository constructed under the set threshold and further evaluated the filtering time of the feature repository and the detection time of TPL. It should be noted that we ignored the loading data time cost in the processing.

\begin{table}[htbp]
\small
\setlength{\tabcolsep}{0.19cm}
\caption{TPL Repository Filtering Time Cost}
\begin{tabular}{|c|ccc|c|}
\hline
 & T1 & T2 & T3 & Total\\
\hline
AVG. Time (seconds) & 1.505 & 150.786 & 550.253 & 702.544 \\
\hline
\end{tabular}
\label{tab:TPL_Repository_Filtering_Time}
\end{table}

\begin{table}[htbp]
\small
\setlength{\tabcolsep}{0.07cm}
\caption{TPL Detection Comparison Time}
\begin{tabular}{|c|cccc|}
\hline
&  BinCoFer & ModX & B2SFinder & LibAM\\
\hline
 AVG. Time (seconds) & \zyy{7.560} & \zyy{2626.413} & \zyy{19.020} & \zyy{1043.063}\\
 
\hline
\end{tabular}
\label{tab:TPL_Detection_Comparison_Time}
\end{table}

\begin{table}[htbp]\zoe
\small
\setlength{\tabcolsep}{0.35cm}
{\zoe}\caption{\zoe{BinCoFer's Performance with Data Volume Increase}}

\begin{tabular}{|c|cccc|}
\hline
data percentage & 25\% & 50\% & 75\% & 100\% \\
\hline
Precision & 0.727 & 0.829 & 0.870 & 0.893\\
Recall & 0.727 & 0.725 & 0.727 & 0.649 \\
F1 & 0.727 & 0.773 & 0.792 & 0.752 \\
\hline
\end{tabular}
\label{tab:Data Volume Performance}
\end{table}

\textbf{TPL feature repository filtering time.}
In the feature filtering stage, we divided the time into three independent parts for calculation purposes: the time to filter the export function (t1), the time to compare the similarity between functions and calculate the frequency of each function to assign weight to functions in a TPL (t2), and the time to calculate the MI complexity (t3). The step-by-step time consumption is shown in Table \ref{tab:TPL_Repository_Filtering_Time}. Finally, we filtered and retained 11,499 functions from 123,867 function features in the TPL feature repository, accounting for 0.092. Each retained function is represented by a 768-dimensional vector and assigned a weight. The total time consumed was 702.544 seconds. 
To avoid wasting computing resources and significant time overhead, we used matrix multiplication for batch similarity calculation during comparison. Therefore, the number of functions to be matched each time was set to 128. The experiments were performed on a Linux environment with an Intel(R) Xeon(R) Gold 6326 CPU @ 2.90GHz with 64 cores and an NVIDIA A100-SXM4-80GB GPU.

\textbf{TPL detection time.}
The time cost of \emph{BinCoFer} in TPL detection can be divided into three steps: data preprocessing, feature extraction, and comparison with the feature repository.
We used the \emph{jTrans} fine-tuned model for feature extraction and matrix methods to detect key TPL functions in the target binary program. Each time, we matched all functions in the binary with the remaining functions in a TPL.

For a fair comparison, we did not account for the time spent on preprocessing the feature library and the target binary, nor did we consider the time taken to load the data. The results of our comparative experiments are shown in Table \ref{tab:TPL_Detection_Comparison_Time}.
In our dataset, the size of the binaries to be detected ranges from 700KB to \zyy{8000KB}. On average, BinCoFer took \zyy{7.560} seconds to detect a binary, while ModX took \zyy{2626.413} seconds, LibAM took \zyy{1043.063} seconds, and B2SFinder took \zyy{19.020} seconds. BinCoFer proved to be the most efficient.
We did not compare the efficiency with BinaryAI because in the online platform, we are unable to eliminate the data loading time.
\zyy{This also demonstrates that when module partitioning and graph matching are not used, the time consumption is reduced, and the approach that first constructing a feature library is more efficient in real-time testing.}
\zyy{We observed the experimental results in Table \ref{tab:Data Volume Performance} of changing the number of test data sets from 25\% of the current test set size, 50\%, 75\%, to 100\%, and found that as the amount of test data increased, precision increased and stabilized, while recall fluctuated and showed a downward trend. This may be due to the fact that when filtering the feature library, a small number of functions that were not screened out due to internal reuse may cause a certain proportion of false alarms. It is expected that precision will be in a stable and relatively high state at a large data volume. The threshold is set strictly, and common functions are screened out in the feature filtering stage, so only when these features in the TPL repository are reused, FNs will occur during detection, resulting in a decrease in recall. This is also due to the fact that binary form TPLs cannot know the creation time of source program functions, and when the source code is unavailable, a function that is duplicated multiple times cannot be found in which it was first created in the TPL, so in scenarios where both binary and TPL source code are unavailable, the final results will inevitably result in FNs.}

\begin{titleEnv}
\textit{\textbf{Answering RQ4:}
Without considering the data loading time, the TPL feature repository filtering time is 702.5 seconds, and the average time for TPL detection on the validation dataset (700KB to \zyy{8000KB}) is \zyy{7.560} seconds without considering the data preprocessing time. Comparative experiments with \emph{ModX}, \emph{LibAM}, and \emph{B2SFinder} show that the time cost of TPL detection is reduced by up to 99.7\%.
\zyy{When the test data set increases, recall tends to decrease and precision tends to stabilize. This is believed to be a phenomenon caused by the second point mentioned in the limitation.}
}
\end{titleEnv}
\section{Discussion}\label{sec:discussion}

\subsection{Differences from OSSFP}
The most similar work to ours is the SCA work OSSFP on source code. 
In terms of target detection, BinCoFer is designed for binary programs, while OSSFP is aimed at open-source programs, resulting in different application scenarios and more challenges for BinCoFer.
First, source code can provide more information, such as the creation time of functions. OSSFP can retain the function with the earliest creation time and remove cloned functions from the feature repository. However, for binaries, since it is impossible to obtain the creation time of functions and it is not accurate to determine the similarity between two functions, it is not possible to remove cloned functions.
Second, since source code can directly use hash matching to determine whether two functions are identical, OSSFP considers that the corresponding TPL will be reused if any function in the repository is matched in the target program.
Third, in binary SCA work, there are few open-source datasets, and manual compilation is needed to collect qualified ground truth datasets. In BinCoFer, static link compilation is adopted to simulate binaries with TPL reuse, resulting in high manual overhead. OSSFP, on the other hand, can collect a large amount of qualified data directly through web scraping.

\subsection{Limitations}\label{sec:Limitations}

\emph{BinCoFer} has the following limitations, suggesting future directions for further work:

First, the cosine similarity between function embeddings can only roughly evaluate the similarity between two functions and cannot accurately represent it. Our method does not directly use the similarity threshold to judge function reuse but still uses the similarity between functions, so a better method is needed to solve this problem.

Second, since we can only obtain the TPLs in binary form and do not have access to related information such as the creation time, we cannot directly filter common functions in the same way we filter simple functions. Instead, we retain functions in a TPL with a high frequency of occurrence within the feature library and assign them a lower weight. However, this approach merely retains these functions and still cannot determine which TPL created them, thus not fundamentally solving the issue \zyy{and can cause FNs.}
Therefore, further exploration is necessary.

Third, we observed during our experiments that IDA PRO sometimes misinterprets the export tables of certain programs, resulting in unparsed information. We manually filtered out some data from the ground truth that was clearly affected by this issue, but due to the large volume of data, there may be omissions.

\zyy{Fourth, we believe that TPL detection and version identification should not be accomplished in the same step. TPL detection aims to identify roughly similar code among a large amount of different code, while version identification aims to detect fine-grained differences among a large amount of identical code. We believe that version detection could be achieved by adding version identification as a subsequent step after the output of our work.}

\zyy{Fifth, our work does not explore highly obfuscated binaries or those compiled with aggressive optimizations. Both of these scenarios are worth specific research in the future.}

\zyy{Sixth}, the setting of thresholds is empirical and can be influenced by the dataset, resulting in poor generalization. However, if experiments are conducted on a larger dataset in the future, this effect is expected to be mitigated.
Although we have manually compiled a dataset, the process is very time-consuming and labor-intensive. How to quickly obtain a larger dataset remains a task worth exploring for the future of binary-to-binary TPL detection.

\zyy{Seventh, when selecting the threshold combination, we base it on the F1 metric, rather than forcing precision to be the highest. However, the experimental results show that BinCoFer is more suitable for scenarios where higher precision is required. How to adapt it to scenarios where higher recall is required is also worth exploring in the future.}

\zyy{Eighth, due to the difficulty of verification on large-scale data and the limitations of the dataset, our work has not yet been applied to larger-scale real-world scenarios for detection, nor has it been verified on other architectures. Those are both worth considering in future work.}
\section{Conclusion}\label{sec:conclusion}

In this paper, we propose a static TPL detection tool for C/C++ programs at the binary level called \emph{BinCoFer}. We construct a feature repository at the function granularity and introduce a three-stage purification strategy to highlight core functions and reduce redundancy in the feature repository.
To address the issue of accurately determining the similarity between two binary functions without directly using a threshold, we first calculate similarity scores at the function granularity between the target binary and the functions in the TPL feature repository. We then use weighted coefficients to aggregate these scores to determine the reuse situation.
Additionally, we manually compiled a new binary SCA dataset \zyy{suitable for the partial TPL reuse detection scenario} and made it available to the community, contributing to the lack of datasets in this field.
\zyy{In the experiment section, we first used \zyy{73\%} of the manually compiled dataset to screen for the threshold using the F1 evaluation metric, and then tested the results on the remaining \zyy{27\%} hold-out set.
\emph{BinCoFer} achieves over 20.0\% higher precision and \zyy{7.0\%} higher F1 than related works on our \zyy{validation} dataset.
We conducted ablation experiments to validate the effectiveness of each step.
In terms of scalability, we observed a decreasing trend in recall and the precision is stable and high as the data volume increases, which we believe may be due to the issue mentioned in the second point of the limitations. Compared to other works, our method constructs a feature library first and does not use module division or graph matching, thereby reducing detection time cost by up to 99.7\%.}

\bibliographystyle{elsarticle}
\bibliography{References}

\end{sloppypar}
\end{document}